\documentclass[english]{article}
\usepackage[T1]{fontenc}
\usepackage[latin9]{inputenc}
\usepackage{geometry}
\usepackage{amstext}
\usepackage{setspace}
\makeatletter

\providecommand{\tabularnewline}{\\}

\makeatother

\usepackage{babel}
\begin{document}

\title{Blind Quantum Computation without Trusted Center}

\author{Shih-Min Hung and Tzonelih Hwang\thanks{Corresponding Author, email: hwangtl@ismail.csie.ncku.edu.tw}}
\maketitle
\begin{abstract}
Blind quantum computation (BQC) protocol allows a client having partially
quantum ability to delegate his quantum computation to a remote quantum
server without leaking any information about the input, the output
and the intended computation. Recently, many BQC protocols have been
proposed with the intention to make the ability of client more classical.
In this paper, we propose two BQC protocols, in which the client does
not have to generate photons, but only has to perform either rotation
or reorder on the received photons. \end{abstract}
\begin{description}
\item [{Keywords:}] Blind Quantum Computation; Quantum Cryptography.
\end{description}

\section{Introduction}

Quantum computation is one based on the principle of quantum mechanics.
Compared with classical computation, it provides the advantage on
calculation speed {[}1{]}. For example, quantum computer can simulate
the property of quantum mechanics, and it is very difficult for classical
computer to do that {[}2{]}. Shor\textquoteright s algorithm {[}3{]}
offers an exponential speedup over the best-known classical solution
for factorizing big integers and solving discrete logarithm problems,
and Grover\textquoteright s algorithm {[}4{]} are also much faster
than the best-known classical search algorithms.

However, realization of quantum computer is still an enormous challenge
now. Although quantum computer appears to be very promising, it still
has a long way to go before it becomes popular. Consequently, in the
near future only a few expensive quantum servers can be accessed by
clients who have to perform quantum computation, but with only a limited
quantum ability. Hence, the clients may have to delegate his problem
to a quantum server without revealing his/her information, including
the input, the output and the intended computation revealed to the
server. Blind quantum computation (BQC) protocol is particularly suitable
to satisfy this requirement.

Based on quantum circuit model, Childs {[}5{]} proposed the first
BQC protocol, where clients need quantum memory and the ability to
perform SWAP gate. Arrighi et al. {[}6{]} also proposed a BQC protocol,
in which the client needs to prepare entanglement states and measure
them. However, these protocols are not universal protocols, in the
sense that they only work on certain classical function, and even
the server can reveal partial information of the client. Broadbent
et al. {[}7{]} then presented the first universal BQC protocol, where
a client does not have any quantum computation ability and quantum
memory except generating rotated single photon and the private information
of the client can be unconditionally secure. After that, many BQC
protocols have been proposed with the intention to make the ability
of client more classical. Li et al. {[}8{]} proposed a triple-server
BQC protocol using entanglement state and Xu et al. {[}9{]} proposed
a single-server BQC protocol based on Li et al.\textquoteright s protocol.
Both Li et al. and Xu et al. claimed that in their protocol the client
only needs to have a quantum channel to receive and resend photons.
However, these protocols have to assume the existence of a trusted
center, which is not practical in reality. Besides, Hung et al. {[}10{]}
pointed out that both Li et al.\textquoteright s and Xu et al.\textquoteright s
protocols are not secure because server can get the information of
the client.

This paper intends to design two secure BQC protocols without trusted
center. The clients in the new BQC protocols only have to perform
either rotation operation or reorder the particles.

The rest of this article is organized as follows. Section 2 reviews
Broadbent et al.\textquoteright s BQC protocol. Section 3 proposes
two BQC protocols. Section 4 analyzes the security of two proposed
protocols. Finally, a concluding remark is given in Section 5.

\section{Review Broadbent et al.\textquoteright s BQC protocol}

Before presenting our protocol, let us briefly review Broadbent et
al.\textquoteright s BQC protocol first. Suppose that a client Alice
with limited quantum capability wants to delegate a quantum problem
to a quantum server Bob on the $m$-qubit graph states corresponding
to the graph $G$ without revealing any information about the input,
the output and the intended computation. The Broadbent et al.\textquoteright s
protocol can be briefly described as follows.
\begin{description}
\item [{Step$\,$1.}] Alice prepares $m$ qubits and sends them to Bob.
The state of each qubit is $\left|\theta_{i}\right\rangle =\left|0\right\rangle +e^{i\theta_{i}}\left|1\right\rangle \left(i=1,2,...,m\right)$,
where $\theta_{i}$ is selected randomly from the set $S=\left\{ k\pi/4|k=0,1,...,7\right\} $.
\item [{Step$\,$2.}] Alice asks Bob to generate a brickwork state according
to the graph $G$ specified by her.
\item [{Step$\,$3.}] According to the graph $G$, Bob produces a brickwork
state $\left|G\left(\theta\right)\right\rangle $ by applying CTRL-Z
gates between the qubits sent from Alice.
\item [{Step$\,$4.}] Alice sends $\delta_{i}=\theta_{i}+\phi_{i}'+r_{i}\pi$
to Bob, where $r_{i}\in\left\{ 0,1\right\} $ is randomly selected
by Alice and $\phi_{i}'$ is a modification of $\phi_{i}$ that depends
on the previous measurement outcomes. Then Bob can measure the $ith$
qubits $\left(i=1,2,...,m\right)$ of $\left|G\left(\theta\right)\right\rangle $.
\item [{Step$\,$5.}] Bob performs a measurement on the $ith$ qubit $\left(i=1,2,...,m\right)$
in the basis $\left\{ \left|\pm\delta_{i}\right\rangle \right\} $
and sends Alice the measurement result.
\item [{Step$\,$6.}] Alice can get the computation output from the measurement
result.
\end{description}
In Broadbent et al.\textquoteright s protocol, Alice needs the ability
of generating rotated single photon. However, the ability of generating
rotated single photon for a client is still considered to be very
difficult now.

\section{Proposed BQC protocols}

This section proposes two BQC protocols. Each reduces the client\textquoteright s
quantum ability to only rotation operation on the particles or to
reorder the particles. The details of these two protocols are described
in Sec. 3.1 and Sec. 3.2 respectively.

\subsection{The first proposed protocol}
\begin{description}
\item [{Step$\,$1.}] Bob prepares $m$ qubits and sends them one-by-one
to Alice. The state of each qubit is $\left|+\right\rangle $.
\item [{Step$\,$2.}] Alice preforms rotation operation $Z_{\theta_{i}}$
on each qubit received from Bob, where $\theta_{i}$ is selected randomly
from the set $S$.
\item [{Step$\,$3.}] Alice sends back this qubit to Bob.
\item [{Step$\,$4.}] Since Bob has the $m$ qubit graph state $\otimes_{i=1}^{m}\left|\theta_{i}\right\rangle \left(i=1,2,...,m\right)$,
and only Alice knows the values of $\theta_{i}$, Alice can run Broadbent
et al.\textquoteright s single-server BQC protocol from Step 2, Bob\textquoteright s
preparation, to delegate the quantum problem to Bob.
\end{description}

\subsection{The second proposed protocol}
\begin{description}
\item [{Step$\,$1.}] Bob generates $m$ Bell pairs $\left|\psi_{0,0}\left(B_{k},A_{k}\right)\right\rangle =\frac{1}{\sqrt{2}}\left(\left|00\right\rangle +\left|11\right\rangle \right)\left(k=1,2,...,2n\right)$
and sends the particle $A_{k}$ of each Bell state to Alice.
\item [{Step$\,$2.}] After Alice receives all particles sent from Bob,
she reorders those particles and sends them back to Bob.
\item [{Step$\,$3.}] Alice sends $m$ classical message $\left\{ \theta_{i}\right\} _{i=1}^{m}$
to Bob, where $\theta_{i}$ is selected randomly form the set $S$.
\item [{Step$\,$4.}] Bob measures his $m$ particles $B_{k}$ in the basis
$\left\{ \pm\theta_{k}\right\} _{k=1}^{m}$ and sends the measurement
results $\left\{ b_{k}\right\} _{k=1}^{m}$ to Alice.
\item [{Step$\,$5.}] Upon receiving $\left\{ b_{k}\right\} _{k=1}^{m}$
form Bob, she knows the state of each $A_{k}$ Bob kept by the measurement
results $\left\{ b_{k}\right\} _{k=1}^{m}$ and the reorder information.
\item [{Step$\,$6.}] Since Bob has the qubit graph state $\otimes_{i=1}^{m}\left|\theta_{i}+b_{i}\pi\right\rangle \left(i=1,2,...,m\right)$,
and only Alice knows the values of $\theta_{i}$ and $b_{i}$, Alice
can run Broadbent et al.\textquoteright s single-server BQC protocolfrom
Step 2, Bob\textquoteright s preparation, to delegate the quantum
problem to Bob.
\end{description}

\section{Security Analysis and Comparison}

\subsection{Security Analysis}

In this section, we discuss the security analysis about both proposed
protocols. Since the security of Broadbent et al.\textquoteright s
BQC protocol has been proved, we only focus on the privacy of $\left\{ \theta_{i}\right\} _{i=1}^{n}$;
if Bob obtains the $\theta$ of each particles, then he can calculate
the input, the output and the intended computation. 

In the first proposed protocol, since Alice performs the rotation
operation $Z_{\theta}$ on the particles in private and then sends
them back to Bob, only Alice knows the $\theta$ of each particle.
Hence, this protocol is as secure as Broadbent et al.\textquoteright s
protocol.

In the second proposed protocol, because Bob knows the measurement
basis $\pm\theta$ and the measurement result for each particle $B_{k}$,
he can calculate the state of $A_{k}$. However, since only Alice
knows the new order of the particles $\left\{ A_{k}\right\} _{k=1}^{m}$
sent from Alice to Bob, Bob cannot find which two particles have entanglement.
Hence, Bob cannot know the state of each $A_{k}$, and hence this
protocol is as secure as Broadbent et al.\textquoteright s protocol.

\subsection{Comparison}

In this sub-section, we give a comparison of Broadbent et al.\textquoteright s
BQC protocol and two proposed BQC protocols (see also Table 1). In
Broadbent et al.\textquoteright s protocol, the client needs the ability
to generate rotated single qubits; in the first proposed protocol,
the client\textquoteright s ability is reduced to preforming rotation
operation; in the second proposed protocol, the client only needs
to reorder the particles. However, to achieve this reduction on the
client\textquoteright s ability, the second proposed protocol pays
the cost: it needs some devices to prevent Trojans horse attack, and
the qubit efficiency of second proposed protocol is lower than Broadbent
et al.\textquoteright s protocol and the first proposed protocol.

\begin{table}
\caption{Comparison table}

\begin{centering}
\begin{tabular}{|c|c|c|c|}
\hline 
 & $\begin{array}{c}
\textrm{Broadbent\text{\textquoteright}s}\\
\textrm{protocol}
\end{array}$  & $\begin{array}{c}
\textrm{The first }\\
\textrm{proposed protocol}
\end{array}$  & $\begin{array}{c}
\textrm{The second }\\
\textrm{proposed protocol}
\end{array}$ \tabularnewline
\hline 
\hline 
Client ability  & generate qubits & rotation operation & reorder\tabularnewline
\hline 
Qubit efficiency & 1/1 & 1/1 & 1/2\tabularnewline
\hline 
Trojans horse attack & $\begin{array}{c}
\textrm{Automatically}\\
\textrm{prevent}
\end{array}$  & $\begin{array}{c}
\textrm{Automatically}\\
\textrm{prevent}
\end{array}$  & $\begin{array}{c}
\textrm{Need device}\\
\textrm{to prevent}
\end{array}$ \tabularnewline
\hline 
\end{tabular}
\par\end{centering}

\end{table}

\section{Conclusions}

This paper has proposed two BQC protocols for a client to delegate
a quantum computation to a remote quantum server without revealing
the input, the output and the intended computation. In both proposed
protocols, the client needs less quantum ability than in Broadbent
et al.\textquoteright s protocol. Whereas the client in the first
protocol needs to preform rotation operation on the particles, the
client in the second protocol only needs the ability to reorder the
particles. We have shown that both proposed protocols are as secure
as Broadbent et al.\textquoteright s protocol. Yet, how to further
reduce a client\textquoteright s quantum ability in a BQC without
revealing any client\textquoteright s information will be a promising
future work.

\section*{Acknowledgment}

We would like to thank the Ministry of Science and Technology of Republic
of China for financial support of this research under Contract No.
MOST 104-2221-E-006-102 -.

\end{document}